\documentclass[%
prd,
superscriptaddress,
prd,
showpacs,showkeys,
 amsmath,amssymb,
 aps,
onecolumn]{revtex4}

\usepackage{changes}
\usepackage{graphicx}% Include figure files
\usepackage[colorlinks=true,citecolor=blue,linkcolor=blue,anchorcolor=green, anchorcolor=green,urlcolor=blue]{hyperref}
\usepackage{mathpazo}
\usepackage{mathtools}
\usepackage{amsfonts}
\usepackage{mathrsfs}
\usepackage{xcolor}
\usepackage{color}
\let\originalleft\left
\let\originalright\right
\renewcommand{\left}{\mathopen{}\mathclose\bgroup\originalleft}
\renewcommand{\right}{\aftergroup\egroup\originalright}
\mathcode`\*="8000
{\catcode`\*=\active\gdef*{\mathclose{}\,\mathopen{}}}

\newcommand{\be}{\begin{equation}}
\newcommand{\ee}{\end{equation}}
\newcommand{\bea}{\setlength\arraycolsep{2pt} \begin{eqnarray}}
\newcommand{\eea}{\end{eqnarray}}

\begin{document}

\title{Weak gravitational lensing by Kerr-MOG Black Hole and Gauss-Bonnet theorem}

\author{Ali \"{O}vg\"{u}n}
\email{ali.ovgun@pucv.cl}
\homepage{https://aovgun.weebly.com} 
\affiliation{Instituto de F\'{\i}sica,
Pontificia Universidad Cat\'olica de Valpara\'{\i}so, Casilla 4950,
Valpara\'{\i}so, Chile.} 
\affiliation{Physics Department, Arts and Sciences Faculty, Eastern Mediterranean
University, Famagusta, North Cyprus via Mersin 10, Turkey.}
%\affiliation{Institute for Advanced Study, 1 Einstein Drive Princeton, N.J. 08540, U.S.A.}

\author{\.{I}zzet Sakall{\i}}
\email{izzet.sakalli@emu.edu.tr} \email{izzet.sakalli@gmail.com}
\affiliation{Physics Department, Arts and Sciences Faculty, Eastern Mediterranean
University, Famagusta, North Cyprus via Mersin 10, Turkey.}

\author{Joel Saavedra}
\email{joel.saavedra@ucv.cl} 
\affiliation{Instituto de F\'{\i}sica, Pontificia Universidad Cat\'olica de
Valpara\'{\i}so, Casilla 4950, Valpara\'{\i}so, Chile.}
\date{\today}

\begin{abstract}
  The deflection angle of Kerr-MOG black holes is studied for different values of the parameter in modified gravity (MOG).  To this end, we employ the Gauss-Bonnet theorem, which was first studied by Gibbons and Werner and then extended by Ono, Ishihara and Asada, who use a generalized optical metric where the deflection of light for an observer and source at finite distance. By using this method, we study the weak gravitational lensing by Kerr-MOG black hole. Our computations show that with an increase in the MOG parameter ($\alpha$), the deflection angle becomes significantly larger than that of Kerr black hole. The results obtained show that MOG effect could be taken into account in the gravitational lensing experiments.
\end{abstract}

\keywords{ Light deflection, Gauss-Bonnet theorem, Gravitational lensing, Black hole}
\pacs{04.40.-b, 95.30.Sf, 98.62.Sb}
\maketitle

\section{Introduction}

In 1783, having accepted Newton's corpuscular theory of light, which conjectured that photons consist of ultra-tiny particles, geologist John Michell was the first scientist known to propose the existence of  dark stars, which are known today as black holes. Michell sent a letter to the \textit{Philosophical Transactions of the Royal Society of London} \cite{schaffer} in which he reasoned that such ultra-tiny particles of light, when emitted by a star, are decelerated by the star gravitational acceleration, and thought that it might therefore be possible to estimate the mass of the star based on the decreasing in their speeds. On the other hand, a star's gravitational attraction might be so strong that the escape velocity could exceed the speed of light. Since the light could not escape from such a star, it would be dark or invisible. Michell estimated that this would be the case of a star more than 500 times the size of the Sun. Michell also claimed that astronomers might detect the dark stars by analyzing star systems gravitationally behaving like binary stars, but where only one star could be observed. Today, astronomers believe that stellar dark stars (black holes) do indeed exist at the centers of many galaxies. Most of the stellar black hole candidates in our galaxy (the Milky Way) are in the X-ray compact binary systems \cite{aitken}. Michell's idea went neglected for more than a century, because it was believed that light could not be interfaced with the gravity. However, in 1915, general relativity theory (GRT) of Einstein revealed that a gravitational lens [a distribution of matter (such as a cluster of galaxies) between a distant light source and an observer] can bend the light from the source as the light travels towards the observer. This effect is known as gravitational lensing a prediction subsequently borne out by experiment \cite{dyson,longair} in 1919. The term black hole was coined by the quantum physicist John Wheeler, who also gave wormholes their name and argued about the nature of reality with Einstein and Bohr \cite{misner}.
Since 1919, which is the year of the experimental verification of the bending of light of GRT, numerous studies on the gravitational lensing have been made not only for the black holes but also for the other astrophysical objects such as wormholes, cosmic strings, global monopoles, neutron stars, etc. \cite{darwin,bozza,iver,bozza2,Bozza:2009yw,fritelli,virbhadra,virbhadra2,virbhadra3,virbhadra4,virbhadra5,zschocke,Stefanov:2010xz,Bartelmann:1999yn}. 

In 2008, Gibbons and Werner (GW) came up with a new idea to compute the deflection angle of light \cite{Gibbons:2008rj}. GW posited that both source and receiver are in the asymptotic Minkowski region. In the sequel, they applied the Gauss-Bonnet theorem (GBT) to a spatial domain, which is defined by the optical metric \cite{Gibbons:2008rj}. In 2012, Werner calculated the deflection angle of Kerr black hole using the optical geometry that has a surface with a Finsler-Randers type with GBT and Naz{\i}m's osculating Riemannian manifold \cite{Werner:2012rc}. Thus, one can evaluate the associated deflection angle integral in an infinite domain bounded by the light ray with GBT \cite{Gibbons:2008hb,Werner:2012rc,Gibbons:2015qja,Gibbons:2011rh,Gibbons:2008zi,Sakalli:2017ewb,Jusufi:2017lsl,Crisnejo:2018uyn,Jusufi:2017hed,Ovgun:2018ran,Jusufi:2017vta,Ovgun:2018prw,Jusufi:2017mav,Ovgun:2019prw,Jusufi:2017xnr,Cvetic:2016bxi,Jusufi:2017uhh,Ovgun:2018fnk,Jusufi:2018jof}. 
In GBT, one can use a domain $\mathcal{D}_{R}$, which is bounded by the light ray as well as a circular boundary curve $C_R$ that is located at center on the lens where intersect the light ray at source and receiver. It is assumed that both source and receiver are at the coordinate distance $R$ from the lens). The GBT is expressed in the optical metric, in the weak field approximation, as follows: 
\begin{equation}
\iint\limits _{\mathcal{D}_{R}}K\,\mathrm{d}S+\oint\limits _{\partial\mathcal{D}_{R}}\kappa\,\mathrm{d}t+\sum_{i}\theta_{i}=2\pi\chi(\mathcal{D}_{R}).
\end{equation}
Note that $K$ is the optical Gaussian curvature and $dS$ is an areal element. After considering the Euler characteristic $\chi(\mathcal{D}_{R}) =1$ and summing the jump angles $\sum_{i}\theta_{i}=\pi$, the deflection angle is obtained by using the following equation acting in compliance with the straight line approximation: \cite{Gibbons:2008rj}

\begin{eqnarray} 
\hat{\alpha} & = & -\int\limits _{0}^{\pi}\int\limits _{r}^{\infty}K \,\mathrm{d}S. \label{Alphaa}
\end{eqnarray}

On the other hand, without considering the asymptotic receiver and source, A. Ishihara et al. \cite{Ishihara:2016vdc,Ishihara:2016sfv} extended GW's method to get the finite-distance corrections to both small and strong deflections. It is worth noting that strong deflection limit is for the light orbits that may have the winding number larger than one \cite{Ishihara:2016sfv}. But, the method of \cite{Ishihara:2016vdc,Ishihara:2016sfv} are limited to the spherical symmetry. Very recently, Ono, Ishihara, and Asada (OIA) \cite{Ono:2017pie,Ono:2018ybw} have generalized the technique of A. Ishihara et al. \cite{Ishihara:2016vdc,Ishihara:2016sfv,Arakida:2017hrm} to the axisymmetric spacetimes that may consist of gravitomagnetic effects. By this way, they have obtained how the the finite-distance corrections effect the deflection angle of light in the axisymmetric spacetime. In the setup of \cite{Arakida:2017hrm}, a photon orbit possesses a non-vanishing geodesic curvature, though the light ray in the 4D spacetime obeys a null geodesic. In particular, they showed that the deflection angle is coordinate-invariant in the framework of the GBT. 

As is well known, Kerr solution \cite{kerr} is one of the best solutions to the Einstein's field equations. Although it is designed to model the astrophysical rotating black holes, recent theoretical and astrophysical studies show that Kerr solution of GR could be modified \cite{Moffat:2006rq,Moffat:2005si,Moffat:2015kva,Moffat:2014aja}. Among the modified gravities (MOGs), the scalar tensor vector gravity theory (STVGT) of Moffat \cite{Moffat:2005si} explains the rotation curves of a galaxy and the dynamics of galactic clusters \cite{Moffat:2006ii,Moffat:2013sja}. The good thing about using the MOG is that it does not need for the dark matter when making those explanations. For this reason, Kerr-MOG black holes have gained more attention in the last years \cite{Moffat:2006gz,Sharif:2017owq,Lee:2017fbq,Guo:2018kis}. Recently, the MOG theory was examined by using the weak gravitational lensing of the Bullet Cluster and merging clusters in Abell 520 \cite{bullet1,bullet2}. The results of this observation provide the validity of the MOG for the Bullet Cluster and show a perfect agreement between weak gravitational lensing and the MOG predictions. Furthermore the theory of the MOG helps to understand the solar system and rotational curves of galaxies \cite{Moffat:2006rq}.

In this paper, our computations about the weak gravitational lensing of Kerr-MOG black hole utilize the method of Gauss-Bonnet first prescribed by Gibbons and Werner \cite{Gibbons:2008rj}, which reveals the ignored role of topology in gravitational lensing. Gibbons and Werner’s seminal work motivated us to study the effect of MOG, which is developed as an alternative to dark matter that opens a new window to the astrophysical applications of the theory, on the gravitational lensing \cite{ Moffat:2013sja}. Our results shall put forward that many features of weak gravitational lensing arise from the topological properties of an underlying optical manifold. Beyond this question of principle, we also show that for weak gravitational lensing of (at least) spherically symmetric distributions, the Gauss-Bonnet approach holds several advantages over common methods such as the thin-lens approximation and direct calculations based on the metric components. Previous studies have benefited from these advantages to study the weak gravitational lensing in various contexts \cite{Moffat:2006gz,Sharif:2017owq,Lee:2017fbq,Guo:2018kis,Javed:2019qyg,Ovgun:2019wej}.
  
This paper is organized as follows. Section II is devoted to the introduction of the Kerr-MOG black hole. In Sect. III, we employ the gravitational lensing formulation of OIA \cite{Ono:2017pie,Ono:2018ybw} to compute the gravitational deflection angle of light from the Kerr-MOG metric. Some plots about the gravitational lensing are also illustrated. Finally, we discuss our results and conclude in Sec. IV.

\section{Kerr-MOG black hole spacetime}

In this section, we summarize the Kerr-MOG black hole, which is a to the STVGT. The action of the STVGT is written as follows: \cite{Moffat:2005si,mynewton}
\begin{eqnarray}
 S=S_{GR}+S_{\phi}+S_{S}+S_{M},
\end{eqnarray}
with
\begin{eqnarray}
S_{GR}&=&\frac{1}{16\pi}\int d^{4}x\sqrt{-g}\frac{R}{G},\\
 S_{\phi}&=&\int d^{4}x\sqrt{-g}\left(-\frac{1}{4}B^{\mu\nu}B_{\mu\nu}
     +\frac{1}{2}\mu^{2}\phi^{\mu}\phi_{\mu}\right),\\
 S_{S}&=&\int d^{4}x\sqrt{-g}\frac{1}{G^{3}}
  \left(\frac{1}{2}g^{\mu\nu}\nabla_{\mu}G\nabla_{\nu}G-V(G)\right)
  +\int d^{4}x\frac{1}{\mu^{2}G}
  \left(\frac{1}{2}g^{\mu\nu}\nabla_{\mu}\mu\nabla_{\nu}\mu-V(\mu)\right),
\end{eqnarray}
where $S_{M}$ is for the matter action, $G(x)$ and $\mu(x)$ are the scalar fields, $\phi^{\mu}$ stands for a massive vector field (Proca type), $\mu$ arises as a result of considering a vector field of non-zero mass, and controls the coupling range, as well as, it represents the variable mass of the vector field. The corresponding potentials are $V(G)$ and $V(\mu)$. Moreover, the tensor field can be written as $B_{\mu\nu}=\partial_{\mu}\phi_{\nu}-\partial_{\nu}\phi_{\mu}$, which satisfies:
\begin{eqnarray}
 &&\nabla_{\nu}B^{\mu\nu}=0,\\
 &&\nabla_{\sigma}B_{\mu\nu}+\nabla_{\mu}B_{\nu\sigma}+\nabla_{\nu}B_{\sigma\mu}=0.
\end{eqnarray}
On the other hand, we can write the energy momentum tensor for the vector field as follows:
\begin{eqnarray}
 T_{\phi\mu\nu}&=&-\frac{1}{4\pi}\left(B_{\mu}^{\;\sigma}B_{\nu\sigma}
      -\frac{1}{4}g_{\mu\nu}B^{\sigma\beta}B_{\sigma\beta}\right).
\end{eqnarray}
Then we simplify the action as
\begin{eqnarray}
 S=\int d^{4}x\sqrt{-g}\left(\frac{R}{16\pi G}-\frac{1}{4}B^{\mu\nu}B_{\mu\nu}\right),
\end{eqnarray}
with the Einstein field equation's
\begin{eqnarray}
 G_{\mu\nu}=-8\pi G T_{\phi\mu\nu},\label{field}
\end{eqnarray}
where $G$ stands for the Newton's gravitational constant [$G=G_{N}(1+\alpha)$] with having a dimensionless parameter$\alpha$. The parameter $\alpha$ determines the strength of the gravitational field and can be used for measuring the deviation of MOG from general relativity. Namely, MOG contains GR in the limit that the scalar field parameter vanishes: $\alpha=0$ \cite{Moffat:2016izzet}.

The line-element of the Kerr-MOG black hole in Boyer-Lindquist coordinate is  given by \cite{Moffat:2014aja,mynewton}
\be
ds^2=-\frac{\Delta \Sigma}{\Xi}dt^2+\frac{\Sigma}{\Delta}dr^2+\Sigma d\theta^2+\frac{\Xi \sin^2 \theta}{\Sigma}(d\phi-\omega dt)^2, \label{metric1}
\ee
where
\begin{subequations}
\bea
\omega=\frac{a(2M_{\alpha}r-G_N^2M_{\alpha}^2\frac{\alpha}{1+\alpha})}{\Xi},\quad
\Delta=r^2-2G_NM_{\alpha}r+a^2+G_N^2M_{\alpha}^2\frac{\alpha}{1+\alpha},\\
\Sigma=r^2+a^2\cos^2\theta,\quad
\Xi=(r^2+a^2)^2-\Delta a^2\sin^2\theta,
\eea
\end{subequations}
and
\be
\label{beta}
M_{\alpha}=(1+\alpha)M.\qquad
\ee
$M$ denotes the mass of the Kerr-MOG black hole and $a$ stands for the spin parameter. The metric (\ref{metric1}) is stationary and axially symmetric. We denote by $\xi$ and $\psi$ the Killing vector fields which are the generators of the corresponding symmetry transformations
\begin{equation}
\xi^\mu_{(t)}= (1,0,0,0),  \quad \textrm{timelike Killing vector field,}
 \end{equation} 
\begin{equation}
	\xi^\mu_{(\phi)}= (0,0,0,1),  \quad \textrm{rotational Killing vector field.}
\end{equation}

The existence of those Killing vectors is very useful when computing the ADM mass, the angular momentum, and the total charge \cite{mywald,mywald2} that define the Kerr-MOG metric (\ref{metric1}). Killing vectors also play an important role in the separability of geodesic and wave equations. In fact, $M_{\alpha}$ is the ADM mass and $\hat{J}=M_{{\alpha}}a$ corresponds to the angular momentum of the Kerr-MOG black hole. %Moreover, there exists a deformation parameter ${\alpha}$, which modifies the gravitational constant as $G=G_N(1+{\alpha})$ \cite{Moffat:2014aja}. 
The gravitational charge of the MOG vector field is given by \cite{Moffat:2014aja}
\be
\label{eq: gravitational charge}
Q=\sqrt{{\alpha} G_N}M.
\ee

From here on in, without loss of generality, we shall adopt the Newton's gravitational constant to unity: $G_N = 1$. The horizons (inner and outer) (at $\Delta=0$) are given by
\be
\label{eq:EventHorizon}
r_{\pm}=M_{{\alpha}}\pm\sqrt{M_{{\alpha}}^2-(a^2+\beta^2)}.
\ee
Note that there is an extremal limit for $a^2+\beta^2=M_{{\alpha}}^2$ where $\beta^2=G_N^2M_{\alpha}^2\frac{\alpha}{1+\alpha}$.

\section{Deflection angle of light by a Kerr-MOG Black hole}

In this section, by using the OIA method, we shall study the deflection angle for the Kerr-MOG black hole. To this end, we consider
non-asymptotic receiver and source. Furthermore, we use a two-dimensional orientable surface with boundaries shown in Fig. 1. It is worth noting that MOG has only one metric and both photons and gravitons travel on null geodesics of this metric, as such in general relativity. Because of this, MOG is not a bimetric or bigravity theory \cite{PLB2018}. Now, we present the orbit equation on the equatorial plane for the spacetime given in Eq. (\ref{metric1}). For this purpose, we first impose the null condition ($ds^2 = 0$) and obtain the orbit equation on the equatorial plane ($\theta=Pi/2$). The timelike and rotational Killing vectors of the Kerr-MOG spacetime (\ref{metric1}) engender geodesic constants of motion identified respectively with energy and the angular momentum: $L=\xi^\mu_{(\phi)} p_\mu$ and $E=-\xi^\mu_{(t)} p_\mu$, which are given by

\begin{align}
E&=\left(\frac{\Delta \Sigma}{\Xi}+\frac{\omega^2 \Xi}{\Sigma}\right)\dot{t}+\left(\frac{\omega \Xi}{2 \Sigma}\right)\dot{\phi} , 
\label{EEE} 
\\
L&=\left(\frac{\Xi}{\Sigma}\right)\dot{\phi}-\left(\frac{\omega \Xi}{2 \Sigma}\right)\dot{t}, 
\label{L}
\end{align}

where dot mark denotes the derivation with respect to the affine parameter. Then the impact parameter $b$ is defined as

\begin{align}
b &\equiv \frac{L}{E} =-{\frac {{\Xi}^{2} \left( \omega\,{\dot{t}}-2\,{\dot{\phi}} \right) }{
 \left( 2\,{\omega}^{2}{\Xi}^{2}+2\,\Delta\,{\Sigma}^{2} \right) {\dot{t}}+{\Xi}^{2}\omega\,{\dot{\phi}}}}. \label{b}
\end{align}

When we use the transformation of $u \equiv 1/r$ and the orbit equation becomes

\begin{align}
\left(\frac{du}{d\phi}\right)^2
= F(u) , 
\label{OE-2}
\end{align}
where $F(u)$ is 
\begin{align}
F(u) 
=-{\frac {{u}^{4}\Xi\, \left( 5\,{\omega}^{2}{\Xi}^{2}+4\,\Delta\,{
\Sigma}^{2} \right)  \left( {\Xi}^{2}{b}^{2}{\omega}^{2}+\Delta\,{
\Sigma}^{2}{b}^{2}+{\Xi}^{2}\omega\,b-{\Xi}^{2} \right) \Delta}{{
\Sigma}^{2} \left( 2\,{\Xi}^{2}b{\omega}^{2}+2\,\Delta\,{\Sigma}^{2}b+
{\Xi}^{2}\omega \right) ^{2}}}. 
\label{F-axial}
\end{align}

Now, we follow the OIA method to compute the deflection angle. Rewriting the null condition $ds^2=0$, we obtain \cite{Ono:2017pie}
\begin{align}
  dt = & \sqrt{\gamma_{ij} dx^i dx^j} + \eta_i dx^i , \label{opt}
\end{align}

in which $\gamma_{ij}$($i, j = 1, 2, 3$)$\neq$$g_{ij}$: 3-dimensional Riemannian space $^{(3)}M$ in which the motion of the photon is described as a motion in a spatial curve. Coordinates of $t$ and $\phi$ are associated with the Killing vectors. In the equatorial plane ($\theta = \frac{\pi}{2}$), after some algebra, one may define the arc-length ($\ell$) along the light ray as \cite{Ono:2017pie}
\begin{align}
d \ell^2= \gamma_{ij} dx^i dx^j = & \left( \frac{\Delta \Sigma}{\Xi} - \frac{\omega^2
  \Xi}{\Sigma} \right) dr^2 + \frac{\Sigma}{\Delta} d \phi^2,
\end{align}

and

\begin{align}
  \eta_i dx^i = & - \omega d \phi .
\end{align}

The unit tangent vector $e^i$ directed along the photon orbit is defined as follows ($\gamma_{ij} e^i e^j = 1$) \cite{Ono:2017pie}: 

\begin{align}
e^i=\frac{1}{\xi} \Big(\frac{dr}{d\phi}, 0, 1 \Big), 
\label{ei}
\end{align}
to find the impact parameter and the photon directions at the receiver and source,
where
\begin{align}
\frac{1}{\xi}=2\,{\frac { \left( 2\,{\Xi}^{2}b{\omega}^{2}+2\,\Delta\,{\Sigma}^{2}b+
{\Xi}^{2}\omega \right)  \left( {\omega}^{2}{\Xi}^{2}+\Delta\,{\Sigma}
^{2} \right) }{{\Xi}^{2} \left( 5\,{\omega}^{2}{\Xi}^{2}+4\,\Delta\,{
\Sigma}^{2} \right) }}. 
\label{xi}
\end{align}
For the outgoing direction of the unit radial vector, we have
\begin{align}
R^i= \Big(\frac{1}{\sqrt{\gamma_{rr}}}, 0, 0 \Big). 
\label{R}
\end{align}
Thus, using the $\cos \Psi \equiv \gamma _ { i j } e ^ { i } R ^ { j }$, one can write the angle from $R^i$ as follows:

\begin{align}
\sin\Psi 
=&\sqrt{1-(\gamma_{i j}e^iR^j)^2}={\frac {2\,{\Xi}^{2}b{\omega}^{2}+2\,\Delta\,{\Sigma}^{2}b+{\Xi}^{2}
\omega}{\Xi\,\sqrt {5\,{\omega}^{2}{\Xi}^{2}+4\,\Delta\,{\Sigma}^{2}}}
}.\label{sin}
\end{align}

The boundary of the integration domain with an impact parameter $b$ now reads

\begin{align}
  \sin \Psi = \frac{\sqrt{\beta^2 + r^2} b}{r^2} - \frac{\beta^2 a}{r^4
  \sqrt{\beta^2 + r^2}} - \frac{bM_{{\alpha}}}{r^{} \sqrt{\beta^2 + r^2}}.
  \label{u}
\end{align}

Thus, the straight line approximation of the light ray, in the weak-approximation and slowly rotating limits, yields
\begin{align}
 u = \frac{\sin \phi}{b} +
\frac{M_{{\alpha}} (1 + \cos^2 \phi)}{b^2} - \frac{2 a M_{{\alpha}}}{b^3}.
\end{align}

Exact deflection angle \cite{Ishihara:2016vdc} can be calculated via
\begin{equation}
\hat{\alpha}=2 \int^{u_0}_0 \frac{du}{\sqrt{F(u)}}-\pi, 
\end{equation}
where $u_0$ is the inverse of the closest approach. One can now express the deflection angle with the new angles of receiver $\Psi_R$,
source $\Psi_S$, and coordinate $\phi_{RS}$ as follows
\begin{equation}
  \hat{\alpha} \equiv \Psi_R - \Psi_S + \phi_{RS} . \label{alpha-axial}
\end{equation}
Note that $\Psi_S$ is the exterior angle at the vertex $S$ and $\Psi_R$ stands for the opposite angle of the interior angle at the vertex $R$.

\begin{figure}[!ht]
\centering
\includegraphics[width=0.4\textwidth]{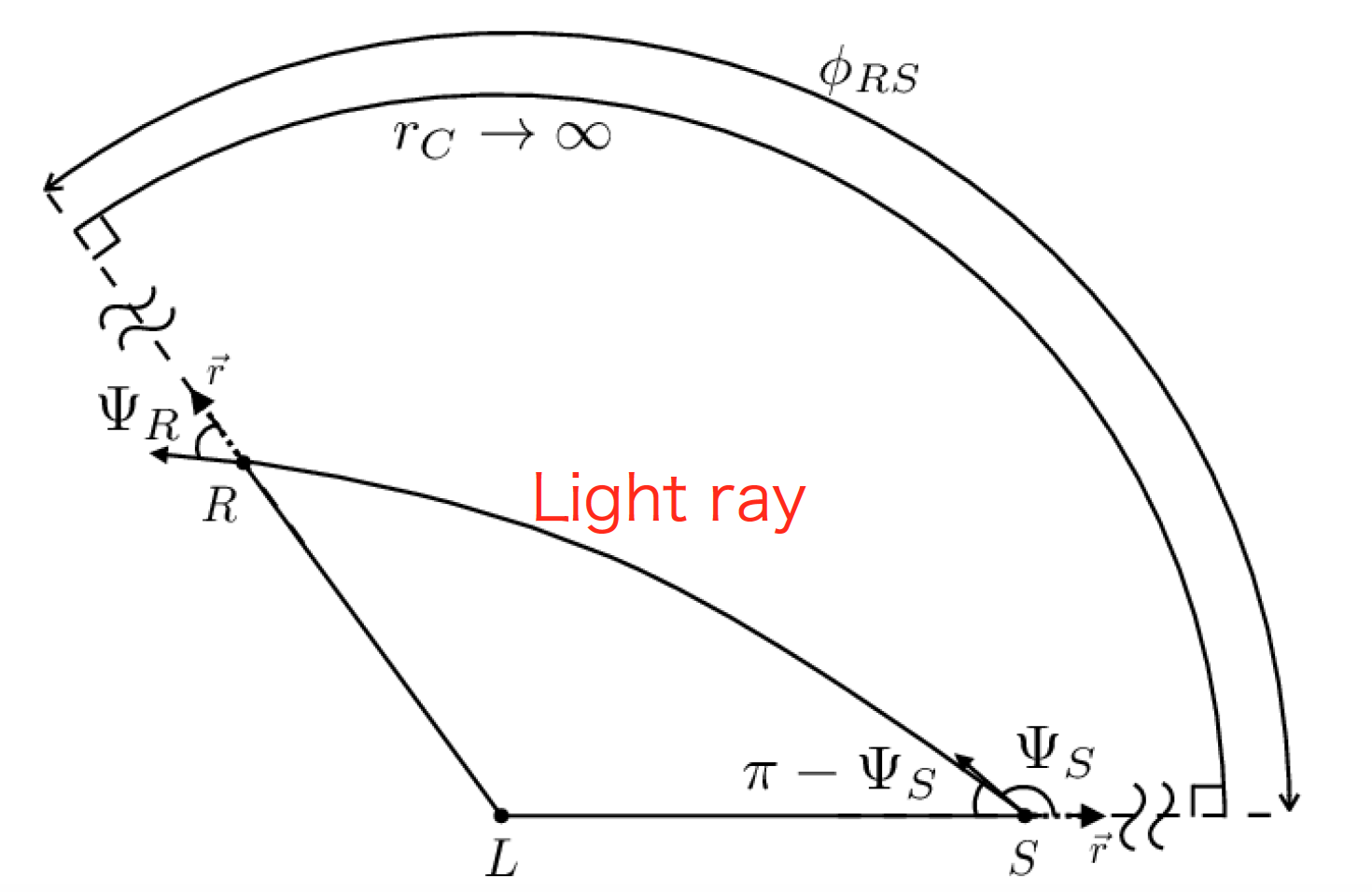}
 \caption{\label{fig:01} Geometrical configuration of quadrilateral embedded in a curved space \cite{Ishihara:2016vdc}.}
\end{figure}

In this approach, we locate the positions of the receiver and source to specific locations. Namely, we require that the endpoints (the receiver and source) of the photon orbit are in Euclidean space in which the angles can be easily determined. Thus, the deflection angle $\hat{\alpha}$ \cite{Ono:2017pie} is obtained as

\begin{align}
  \hat{\alpha} = - \iint_{^{\infty}_R \square^{\infty}_S} KdS + \int_S^R \kappa_g d
  \ell, \label{GB-axial}
\end{align}

where $^{\infty}_R \square^{\infty}_S$ stands for a quadrilateral embedded and
$\kappa_g$ is the geodesic curvature. Taking cognizance of Ref. \cite{Arakida:2017hrm}, one can see that the geodesic curvature $\kappa_g$ is defined as

\begin{align}
  \kappa_g = - \sqrt{\frac{1}{\gamma \gamma^{\theta \theta}}} \beta_{\phi, r}
  . \label{kappa}
\end{align}

For the Kerr-MOG black hole, Eq. (\ref{kappa}) becomes
\begin{align}
  \kappa_g = & - \frac{2 a (1 + {\alpha}) M}{r^3}.
\end{align}

Note that the main contribution of rotation parameter $a$ to the deflection angle comes only from the geodesics [from the source ($\textit{S}$) to the receiver ($\textit{R}$)]. Thus, the deflection angle $\alpha$ is found to be

\begin{align}
 \int_S^R \kappa_g d \ell = & \int^S_{~ R} \frac{2 a (1 + {\alpha}) M}{r^3} d
  \ell= - \frac{2a(1 + {\alpha})M}{b^2}\int^{\phi_R}_{\phi_S} \cos\vartheta d\vartheta 
\notag\\
=&- \frac{2aM}{b^2}[\sqrt{1-b^2{u_R}^2}+\sqrt{1-b^2{u_S}^2}]. \label{intkappa}
\end{align}
Note that during those computations the linear approximation of the photon orbit ($r= b/\cos\vartheta$ 
and $\ell = b \tan\vartheta$) is considered.

At far limit, $u_R \to 0$ and $u_S \to 0$, we have
\begin{equation}
  \int_S^R \kappa_g d \ell =\frac{4 a (1 + {\alpha}) M}{b^2}.
\end{equation}

To evaluate the first integral in Eq. (\ref{GB-axial}), we use the Gaussian optical curvature in
the weak field approximation:

\begin{align}
  K = & \frac{R_{r \phi r \phi}}{\det \gamma}
  = \frac{1}{\sqrt{\det \gamma}}  \left[ \frac{\partial}{\partial \phi} 
  \left( \frac{\sqrt{\det \gamma}}{\gamma_{rr}} \Gamma^{\phi}_{~ rr} \right) -
  \frac{\partial}{\partial r}  \left( \frac{\sqrt{\det \gamma}}{\gamma_{rr}}
  \Gamma^{\phi}_{~ r \phi} \right) \right]\\
  = & - \frac{2 (1+\alpha)M}{r^3} %- \frac{3 (2 M_{{\alpha}} - r) \beta^2}{r^5} 
  +\mathcal{O} (\frac{M^2}{r^4}) .\end{align}

It is worth noting that the contribution of rotating parameter $a$ does not appear in the Gaussian optical curvature part. Namely, the original GBT \cite{Gibbons:2008rj} does not reveal the effect of the rotating parameter $a$ on the deflection angle. That is why we have preferred to use the OIA method in the present study.

Evaluating the Gaussian curvature integral for the Kerr-MOG black hole
in the limit of \ $r_R \to \infty$ and $r_S \to \infty$:

  \begin{align}
-\iint_{{}^{\infty}_{R}\Box^{\infty}_{S}} K dS  
=&
\int^{\infty}_{r_{OE}} 
dr
\int_{\phi_S}^{\phi_R} d\phi   \left[
\frac{2(1+\alpha)M}{r^2} 
+ \mathcal{O} \left(\frac{M^2}{b^2}\right) \right] 
\notag\\
=&
2(1+\alpha)M \int_{\phi_S}^{\phi_R} d\phi  
\int_{0}^{\frac{1}{b}\sin\phi+\frac{(1+\alpha)M}{b^2}(1+\cos^2\phi)
-\frac{2a(1+\alpha)M}{b^3}} du
+ \mathcal{O}\left(\frac{M^2}{b^2}\right) 
\notag\\
=&
2(1+\alpha)M \int_{\phi_S}^{\phi_R} d\phi 
\Big[u\Big]^{\frac{1}{b}\sin\phi+\frac{(1+\alpha)M}{b^2}(1+\cos^2\phi)
-\frac{2a(1+\alpha)M}{b^3}}_{u=0}  
+ \mathcal{O}\left(\frac{M_\alpha^2}{b^2}\right) 
\notag\\
=&
\frac{2(1+\alpha)M}{b} 
\int_{\phi_S}^{\phi_R} d\phi 
\sin\phi  
+ \mathcal{O} \left(\frac{M^2}{b^2}, \frac{aM^2}{b^3}\right) 
\notag\\
=&\frac{2(1+\alpha)M}{b}\Big[\sqrt{1-b^2{u_S}^2}+\sqrt{1-b^2{u_R}^2}\Big] 
+ \mathcal{O}\left(\frac{M^2}{b^2}, \frac{aM^2}{b^3}\right), 
 \label{intK}
\end{align}

at far limit $u_R \to 0$ and $u_S \to 0$, we get
\begin{equation}
-\iint_{{}^{\infty}_{R}\Box^{\infty}_{S}} K dS  
= \frac{4 M (1 + \alpha)}{b} +\mathcal{O} (\frac{M^2_{{\alpha}}}{b^2}).
  \end{equation}
  \begin{figure}[!ht]
\centering
\includegraphics[width=0.6\textwidth]{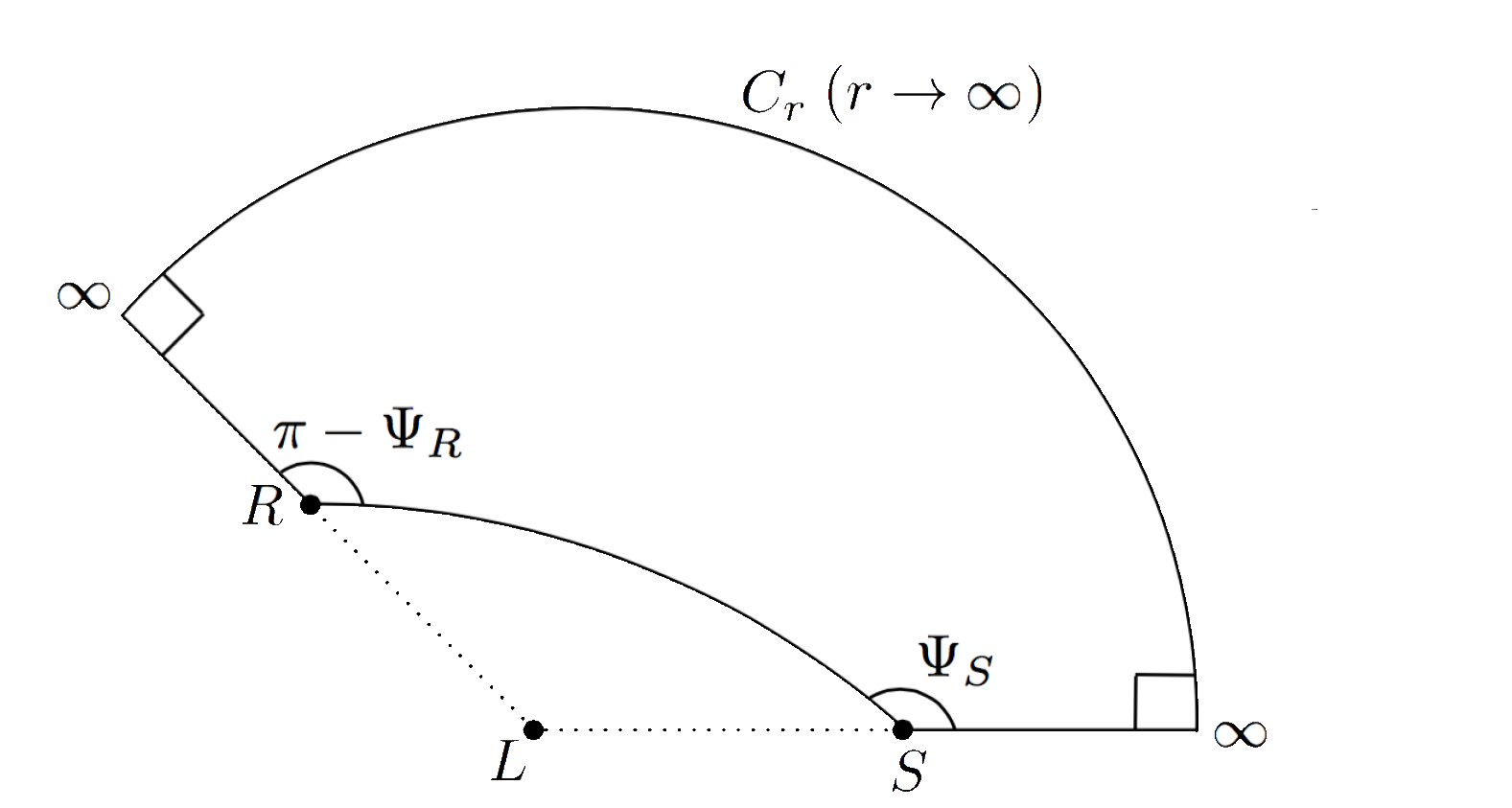}
 \caption{\label{fig:02} Asmptotically flat spacetime \cite{Ishihara:2016vdc}.}
\end{figure}

\begin{figure}[h]
\centering
\includegraphics[width=0.4\textwidth]{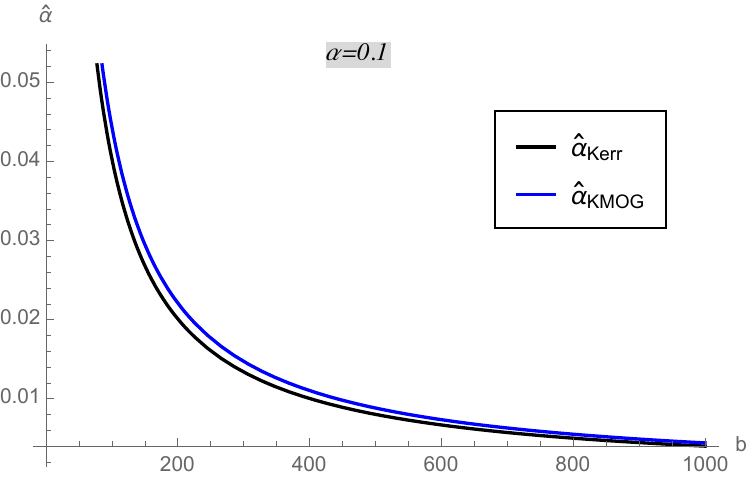}
\includegraphics[width=0.4\textwidth]{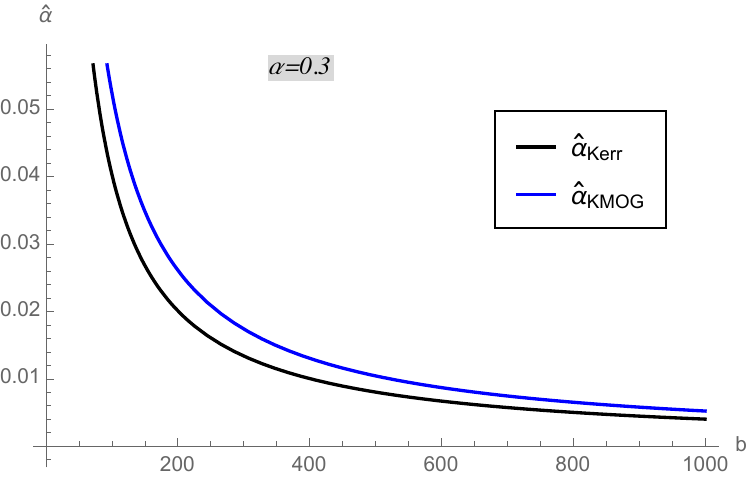}
\includegraphics[width=0.4\textwidth]{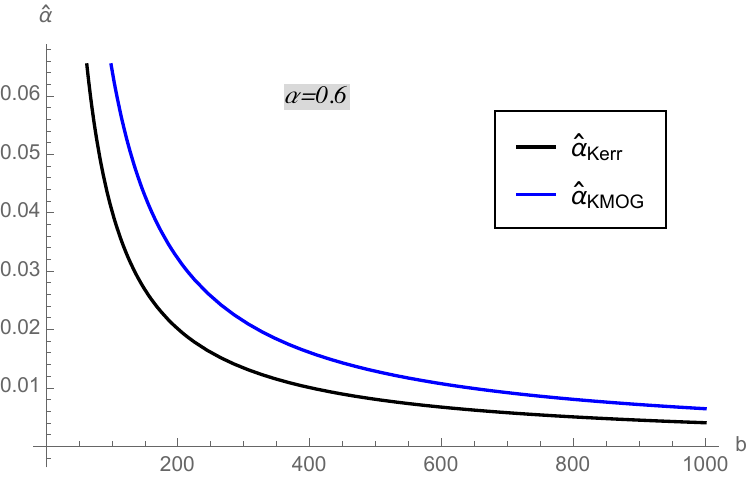}
\caption{\label{fig:1} The plots of deflection angle versus impact parameter $b$ for the Kerr and Kerr-MOG black holes. The physical parameters are chosen as $M=a=1$. Each plot is drawn according to $\alpha=0.1$,  $\alpha=0.3$, and $\alpha=0.6$.} 
\end{figure}

We remark that the finite-distance corrections due to gravitomagnetism are different from the general relativity case. However in the limit of \ $r_R \to \infty$ and $r_S \to \infty$, the results reduce to the general relativity's ones \cite{Ono:2017pie}. After combining the contributions coming from the Gaussian optical curvature and the geodesic
curvature parts, the total deflection angle is obtained as follows:
\begin{equation} \hat{\alpha} = \frac{4 M(1 + \alpha)}{b} \pm \frac{4 a (1 + \alpha) M}{b^2},
\end{equation}
where the positive sign stands for the retrograde and the negative sign is for prograde case of the photon orbit. For the case of $\alpha=0$, it reduces to Kerr case, and for the case of $a=0$ \cite{Moffat:2014aja} it corresponds to the deflection angle of the Schwarzschild black hole in the MOG: see Fig. \ref{fig:1}, which is nothing but the recovery of \cite{Moffat:2008gi}.

\section{Conclusion}

In the weak field approximation, we have studied the gravitational lensing of the Kerr-MOG black hole. To this end, we have followed the method of OIA instead of the GW's GBT method \cite{Gibbons:2008rj} (the Nazim's osculating Riemannian approach using Finsler-Randers metric) to calculate the deflection angle at the leading order of the weak field approximation in which the receiver and source are located at the null infinity. By this way, we have managed to reveal the effects of MOG and rotation parameters on the deflection angle. Our findings are graphically exhibited. It has been shown that the deflection angle of the Kerr-MOG black hole is significantly greater than the original Kerr black hole. We believe that the results we obtained will contribute to the MOG theory, which has been recently tested by using the weak gravitational lensing of the Bullet Cluster and merging clusters in Abell 520 \cite{bullet1,bullet2}. The results of this observation support the validity of the MOG for the Bullet Cluster and show a perfect agreement between weak gravitational lensing and the MOG predictions. Moreover, the theory of MOG has not need for dark matter in the present universe and be consistent with astrophysical and cosmological data and the neutron star merger data \cite{qnmmog1,qnmmog2}. 

In conclusion, we have shown that the MOG effect must be taken into consideration for the astrophysical observations to be made about the black holes. %In the near future, we plan to explore in detail how the MOG parameter $\alpha$ together with the generalized uncertainty principle has an impact onto the sparsity of Hawking radiation \cite{newprd}.

\acknowledgements
The authors are grateful to the Editor and anonymous Referee for their valuable comments and suggestions to improve the paper.
This work is supported by Comisi\'on Nacional
de Ciencias y Tecnolog\'ia of Chile (CONICYT) through FONDECYT Grant N$^{\textup{o}}$ 3170035 (A. \"{O}.). A. \"{O}. is  grateful to Institute for Advanced Study, Princeton for hospitality.

%\section*{References}
%\bibliographystyle{utphys}
%\bibliography{note}
\providecommand{\href}[2]{#2}\begingroup\raggedright

\end{document}